\documentclass{PoS}

\newcommand{\Et}{\mbox{$E_T$}}

\newcommand{\met}{\mbox{$\protect \raisebox{0.3ex}{$\not$}\Et$}}

\title{Top quark pair production at the Tevatron}

\ShortTitle{Top quark pair production at the Tevatron}

\author{\speaker{Fabrizio Margaroli}\thanks{on behalf of the CDF and D0 collaborations}\\
        Purdue University\\
        E-mail: \email{margarol@fnal.gov}}

\abstract{The top quark has been discovered in 1995 by the CDF and D0 experiments located at the two beam-crossing points at the Tevatron $p \bar p$ collider. The top quark is the most massive of the known elementary particles. At hadron-hadron colliders, top quarks are mostly produced in pairs through strong interactions. The measurement of the cross section for top quark pair production ($\sigma_{t \bar t}$) is a test of our understanding of QCD prediction at next-to-leading order (NLO). This measurement requires a deep understanding of the backgrounds to top quark production, thus setting the ground for the measurements of top quark intrinsic properties. Finally, measuring $\sigma_{t \bar t}$ allows to set limits on new physics models that predict new particles coupling to the most recently discovered quark. The top quark discovery needed only few tens of pb$^{-1}$.  The Tevatron integrated luminosity is now about 9\,fb$^{-1}$, more than two orders of magnitude larger. In this document I will present some of the latest measurements of total and differential cross sections for top quark pair production performed by the CDF and D0 collaborations, that analyze up to about 6\,fb$^{-1}$ to provide a refined understanding of this extremely interesting particle. The identification of hadronic jets with substructure is thought to be a crucial tool towards the search for new phenomena at the Large Hadron Collider (LHC) proton-proton collisions at $\ge 7$\,TeV using the CMS and ATLAS detectors. The first search for highly boosted top quarks has been performed by CDF and is presented here.}

\FullConference{35th International Conference of High Energy Physics - ICHEP2010,\\
		July 22-28, 2010\\
		Paris France}

\begin{document}

\section{Introduction}

The top quark has been intriguing physicists ever since its discovery in 1995 at the CDF and D0 experiments\,\cite{discovery}. In fact, its mass was found to be surprisingly large, some thirty times larger than the one of the bottom quark. Having a Yukawa coupling very close to one, speculation arose on whether the top quark might have a special role in the electroweak symmetry breaking mechanism. The Standard Model (SM) predicts that it is produced mostly in pairs at hadron colliders through QCD processes, with its cross section at the Tevatron being calculated up to the (N)NLO\,\cite{theory}. Theoretical and experimental cross sections are quoted in this paper for an assumed top quark pole mass of 172.5\,GeV/c$^2$, compatible with its current best experimental determination\,\cite{topmass}. Several authors agree on the cross section at $p \bar p$ collisions at 1.96\,TeV to be $\sim 7.5$\,pb. With an integrated luminosity soon reaching 10\,fb$^{-1}$,
the number of top pair events produced in the two collisions points where CDF and D0 are located is about 150\,000. The top quark also has a "special" phenomenology in that it is the sole quark that decays before hadronizing. It decays almost 100\% of the times into $W b$, with subsequent $W$ decays to leptons or quarks. Top quark pair decay modes are labeled as lepton+jets, dileptonic, or all hadronic depending on the combination of hadronic or leptonic decays of the two $W$ bosons. The golden mode for top quarks is the lepton+jets signature, where the charged lepton is an electron or muon; the dilepton and all-hadronic channels provide unique challenges but also complementary information. Summing up the event selection acceptance in the three signatures, the CDF and D0 experiments analyze about 9\% of the produced top quark pair events. Leptonic tau decays are identified through their electron or muon decays. Events with hadronically decaying taus are collected by CDF and D0 with different strategies: D0 adopts explicit tau identification to suppress the otherwise large QCD backgrounds where jets mimic both the tau and the neutrino signature, while CDF uses the \met \, plus $b$-jets signature (vetoing identified electrons and muons). This brings the total acceptance up to about 13\%, i.e. the ability to analyze $\sim$1000 top pair events per integrated fb$^{-1}$.

\section{Measurement of total cross sections}

\begin{figure} 
\includegraphics[width=.53\textwidth]{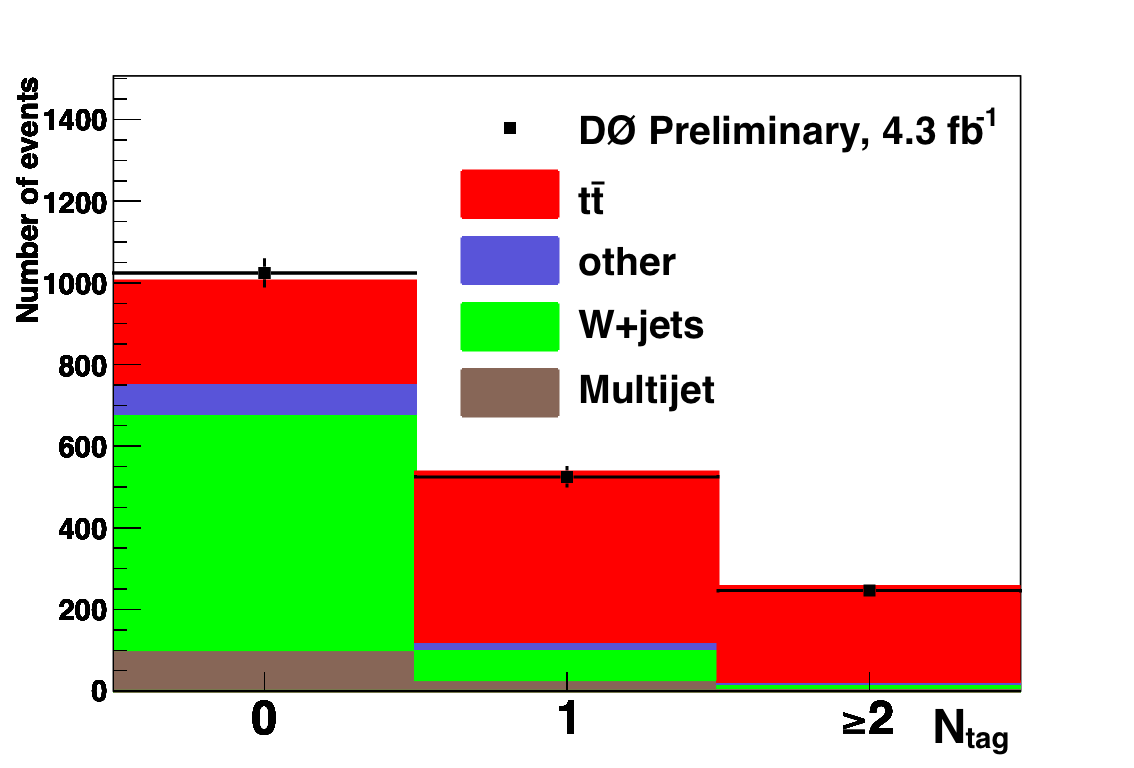} 
\includegraphics[width=.45\textwidth]{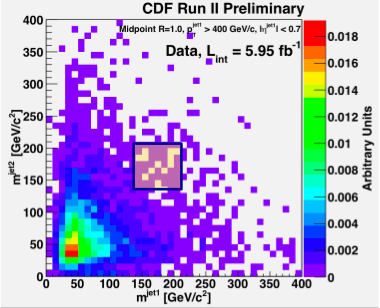} 
\caption{The left plot shows the lepton plus jets sample composition, as a function of the number of $b$-tagged jets. 
The right plot shows the distribution of the two leading jet mass in events selected with one jets with $P_T$>400\,GeV/c. Events with two top quarks decaying hadronically would appear in the boxed region with both jets with mass $\sim M_{top}$.
}
\label{fig1} 
\end{figure}

The most precise measurements come from the lepton+jets signature: both CDF and D0 require the presence of a high $P_T$ electron or muon, large missing transverse energy ($\met$) and at least three high $P_T$ jets. The main background is $W$ plus jets production, where its cross section is poorly known. CDF and D0\,\cite{ljets} use machine-learning techniques using as inputs several kinematical and topological distributions to discriminate between the top quark signal and the $W$ plus jets background. A maximum likelihood fit of the discriminant output distribution allows the extraction of the number of top quark events, and thus of $\sigma_{t \bar t}$. An alternative analysis uses $b$-jet identification algorithms ($b$-tagging) to achieve large signal-to-background ratio (S/B), thus enabling to perform a simpler counting experiment to measure $\sigma_{t \bar t}$. The event yields as a function of the number of $b$-tagged jets can be seen in Fig.\,\ref{fig1} for the D0 analysis. CDF measures the ratio of $t \bar t$ to  $Z \to \ell \ell$, thus effectively trading the dominant luminosity uncertainty for the smaller theoretical uncertainty on the $Z$ production cross section. With a precision below 7\%, this is the most precise determination of $\sigma_{t \bar t}$ as of today. A different CDF analysis uses the very loose signature of $\ge 1$jet, \met and one electron/muon to measure simultaneously the top quark pair signal and the backgrounds contribution\,\cite{CDFshyft}. CDF and D0 measure $\sigma_{t \bar t}$ with lower precision in the dilepton\,\cite{dil} and all-hadronic sample\,\cite{allhad}, where the cause for the former is the limited statistics, and for the latter the poor knowledge of the QCD background.
SM extensions such as supersymmetry (SUSY) predicts the presence of charged Higgs bosons, in addition to the SM neutral one. In this scenario the $t \to H^- b$ decay can compete with the SM decay. Depending on the model parameters, the $H^{\pm} \to \tau^{\pm} \nu$ or  $H^{\pm} \to c \bar s$ decay dominates, thus altering the lepton ratio in top pair decays. D0 measures this ratio for a Higgs mass range $M_W < M_H^{\pm} < M_{top}$, and excludes $BR(t \to H b)$ of 0.1(0.45) at 95\% confidence level (C.L.) for leptonic (hadronic) decays\,\cite{D0Higgs}.
Top quark pair production is a large background to the low mass SM Higgs, SUSY and technicolor signatures, in the \met plus two $b$-jets signature. CDF uses neural networks first to suppress the dominant QCD background in the sample, then to isolate the signal and measure $ \sigma_{t \bar t}$\,\cite{CDFmetjets2}. 
All the CDF and D0 measurements of total $t \bar t$ cross sections are in good agreement with the SM expectations. The two collaborations plan to combine these measurements to further increase precision.

A further stringent test of QCD consists in measuring the $t \bar t + 1$\,jet cross section. This is important also as at the LHC top quark pairs are produced mostly together with extra jet radiation. CDF measures simultaneously $\sigma_{t \bar t + 0 \mathrm{jet}}$ and $\sigma_{t \bar t + 1 \mathrm{jet}}$ in the lepton+jet decay mode, the latter with a precision better than 15\%\,\cite{CDFttplusjet}.

\section{Measurement of differential cross sections}

The measurement of the differential cross section is potentially sensitive to broad enhancements of the spectrum and interference effects.
It has also been suggested that the measurement of $d\sigma_{t \bar t}/dm(t \bar t)$ would allow a measurement of the top quark pole mass $M_{top}$, without the ambiguity of the definition of the top quark mass intrinsic in the Monte Carlo generators.
The cross section $d\sigma_{t \bar t}/d(X)$ where $X$ is either $m(t \bar t)$ or $P_T^{top}$ has been computed at NLO in perturbative QCD. 
CDF measured $d\sigma_{t \bar t}/dm(t \bar t)$ while D0 measured $d\sigma_{t \bar t}/dP_T^{top}$, finding good agreement with the SM predictions\,\cite{diff}.


\section{Search for boosted top quarks}

Top quarks decaying hadronically would appear as a jet with substructure when the quark is very highly boosted. At the Tevatron, this condition is satisfied when the top quark $P_T$ is $> 400$\,GeV/c, with $\sigma_{t \bar t}^{\mathrm{boost}}$ of the order of few fb.  Unfortunately, QCD dijet production still dominates the sample. CDF studied very high $P_T$ jets and isolates the region where at least one top would appear as a jet with mass $\simeq M_{top}$ (see Fig.\,1) and sets limits on $\sigma_{t \bar t}^{\mathrm{boost}} < 55$\,fb at 95\% C.L.

\section{Conclusions}

Top quark physics is a crucial part of the Tevatron program. The measurement of the pair production cross section allows the understanding of the sample composition, fundamental to perform top properties measurements such as its mass, spin, charge, and more. It also allows precision tests of pQCD, and
establishes the $t \bar t$ background to both SM and new physics searches.
The first measurements of differential $t \bar t$ cross sections have been performed with 1\,fb$^{-1}$ of data. Searching boosted top quarks allows studies of jets substructure and the development of tools for searching the Higgs boson and new physics at the LHC. So far, the study of top quark production and decay confirms the SM nature of the top quark. The additional collisions that could be obtained with the proposed Tevatron run extension, and the ones already coming from the LHC collider, will provide by 2014 to the high energy physics community a sample of $O(10^6)$ top quarks, thus further expanding our understanding of the last discovered quark.

\end{document}